\pdfoutput=1 
\documentclass[english,11pt,a4paper]{article}

\usepackage{jheppub} 
\usepackage{graphicx}
\usepackage{bm}
\usepackage{amsmath}
\usepackage{lineno}
\usepackage{amssymb}
\usepackage[utf8]{inputenc}
\usepackage{lipsum}

\newcommand{\be}{\begin{equation}}
\newcommand{\ee}{\end{equation}}

\DeclareUnicodeCharacter{0229}{\c{c}}

\renewcommand\b{\beta}
\renewcommand\d{\delta}

\renewcommand\l{\lambda}

\newcommand\e{\ve}

\newcommand\z{\zeta}

\newcommand\ve{\varepsilon}

\renewcommand\O{\Omega}

\newcommand{\eq}[1]{Eq.~(\ref{#1})}

\newcommand{\eqs}[2]{Eqs.~(\ref{#1})-(\ref{#2})}

\newcommand\lb{\left(}
\newcommand\rb{\right)}
\newcommand\ls{\left[}
\newcommand\rs{\right]}

\newcommand{\non}{\nonumber\\}
\newcommand\pt{\partial}

\newcommand{\bx}{{\bm x}}

\newcommand{\bp}{{\bm p}}
\newcommand{\bk}{{\bm k}}

\newcommand{\bO}{{\bm \O}}
\newcommand{\ba}{{\bm a}}
\newcommand{\bA}{{\bm A}}

\newcommand{\by}{{\bm y}}
\newcommand{\bb}{{\bm b}}
\newcommand{\bB}{{\bm B}}

\renewcommand{\vec}{\boldsymbol}
\renewcommand{\part}{{\rm part}}

\begin{document}

\title{Chiral Vortical Effect For An Arbitrary Spin}

\author[a,b]{Xu-Guang Huang}
\author[c]{Andrey V. Sadofyev}

\affiliation[a]{Physics Department and Center for Particle Physics and Field Theory, Fudan University, Shanghai 200433, China}
\affiliation[b]{Key Laboratory of Nuclear Physics and Ion-beam Application (MOE), Fudan University, Shanghai 200433, China}
\affiliation[c]{Theoretical Division, MS B283, Los Alamos National Laboratory, Los Alamos, NM 87545, USA}

\emailAdd{huangxuguang@fudan.edu.cn}
\emailAdd{sadofyev@lanl.gov}

\abstract{ 
The spin Hall effect of light attracted enormous attention in the literature due to the ongoing progress in developing of new optically active materials and metamaterials with non-trivial spin-orbit interaction. Recently, it was shown that rotating fermionic systems with relativistic massless spectrum may exhibit a 3-dimensional analogue of the spin Hall current -- the chiral vortical effect (CVE). Here we show that CVE is a general feature of massless particles with an arbitrary spin. We derive the semi-classical equations of motion in rotating frame from the first principles and show how by coordinate transformation in the phase space it can be brought to the intuitive form proposed in \cite{Stephanov:2012ki}. Our finding clarifies the superficial discrepancies in different formulations of the chiral kinetic theory for rotating systems. We then generalize the chiral kinetic theory, originally introduced for fermions, to an arbitrary spin and study chirality current in a general rotating chiral medium. We stress that the higher-spin realizations of CVE can be in principle observed in various setups including table-top experiments on quantum optics.
}

\maketitle

\section{Introduction}
Topological effects in optical systems attracted significant attention in the literature, see \cite{Bliokh:2009ek,Bliokh:2015yhi}. One prominent example is the spin Hall effect of light (also known as the optical Magnus effect) \cite{Liberman:1992zz,Bliokh:2004gz,Onoda:2004zz,Duval:2005ky} which corresponds to a separation of circularly polarized photons in the direction transverse to their motion in an optically active medium. This effect originates in a topological phase for photons, often referred to as a Berry phase, in full similarity with its fermionic cousin. Notably, it can be seen as a modification of the velocity along a semi-classical trajectory due to the quantum corrections:
\begin{eqnarray}
\dot{{\bm x}}= \hat {\bm p}+\hbar{\bm F}\times  {\bm b}~~,~~{\bm b}=\pm\frac{\hat {\bm p}}{|{\bm p}|^2}\,,
\label{spinHall}
\end{eqnarray}
where $\hat {\bm p}$ is a unit vector along the momentum direction, ${\bm b}$ is an emergent ``magnetic'' field in the momentum space corresponding to the Berry phase, ${\bm F}$ is an external force, $\hbar$ is an effective Planck constant which is proportional to the wavelength in vacuum. Two signs in ${\bm b}$ correspond to two photon polarizations. In an optically active medium, varying refractive index $n(\bx)$ results in ${\bm F}\sim {\bm \nabla}\,n$ and, thus, light rays of different polarizations propagate with an opposite deflection according to Eq.~(\ref{spinHall}). Further, integrating the velocity with the distribution function, one can arrive at the polarization current in a thermal radiation.

Recently, it was shown that fermionic systems with relativistic spectrum may exhibit bulk topological transports known as chiral effects, see e.g. \cite{Kharzeev:2015znc, Huang:2015oca,Hattori:2016emy}. These phenomena are closely tied with the axial anomaly in relativistic field theories \cite{Adler:1969gk, Bell:1969ts} which, in turn, is related to the Berry phase in the kinetic description \cite{Stephanov:2012ki, Son:2012wh, Son:2012zy, Chen:2012ca, Dwivedi:2013dea, Stone:2013sga, Chen:2014cla,Hidaka:2016yjf,Mueller:2017arw,Huang:2018wdl}. The common origin of spin Hall effect and chiral effects in the fermionic case makes one to expect that spin Hall effect for higher spins may arise along with some bulk topological transports. Such an example of chiral effects for higher spin fields was proposed in \cite{Avkhadiev:2017fxj}. In particular, the authors derived an analogue of the chiral vortical effect (CVE) for vector bosons (photons) which corresponds to a polarization current in a rotating thermal medium along the angular velocity. Soon after this photonic transport was also shown to originate in the Berry phase at the semi-classical level \cite{Yamamoto:2017uul} completing the analogy.

The chiral effects may play a significant role in various systems from quark-gluon plasma (QGP) and primordial plasma in the early Universe to cold atoms and condensed matter systems at lower energies \cite{Kharzeev:2015znc, Joyce:1997uy, Boyarsky:2011uy, Huang:2015mga}. Its relation with the axial anomaly and topological properties of the field configuration makes it a desirable object for an experimental study.  At the moment, some examples of chiral effects were experimentally observed in Dirac- and Weyl-semimetals \cite{Li:2014bha, Gooth:2017mbd} and there is an ongoing search in heavy-ion collisions, see e.g. \cite{Kharzeev:2015znc,Huang:2015oca} and references therein. In this light, an extension of chiral effects to non-fermionic systems looks as a promising direction for further study. First, topologically non-trivial systems of condensed matter, cold atoms and quantum optics, are under active experimental investigation and novel effects can be immediately approached. Second, optical systems can serve as a playground to study many quantum effects in a controlled way and chiral effects for photons may give a guidance for the further considerations of the chiral effects in other systems. It can be also possible to relate the CVE of light with corresponding counterpart of the quantum friction of rotating objects \cite{quantum1, quantum2}. Finally, one may hope to incorporate all parity-odd effects into a closed complete picture which would be important for cosmological and heavy-ion applications, see e.g. \cite{Boyarsky:2011uy,Khaidukov:2013sja, Avdoshkin:2014gpa, Yamamoto:2015gzz, Hirono:2015rla, Hirono:2016jps, Tuchin:2016qww, Avkhadiev:2017fxj, Kirilin:2017tdh, Li:2017jwv, Hattori:2017usa, Tuchin:2018rrw}.

Here we consider the semi-classical description of particles with spin $s\geq\frac{1}{2}$ in a rotating system and extend the fermionic chiral kinetic theory \cite{Stephanov:2012ki, Son:2012wh, Son:2012zy, Chen:2014cla} to this general case. First, we follow \cite{Stephanov:2012ki, Yamamoto:2017uul} including the effect of the spin-orbit coupling and of the Coriolis force intuitively and deriving both the spin Hall effect and the CVE current for arbitrary $s$. It is interesting that within this approach a single-particle/ray picture, commonly used in the literature on the spin Hall effect, can be extended to the CVE case which is to say just modifies classical trajectories. Then, we focus on an explicit derivation of the semi-classical equations of motion (EOMs) for massless particles in a rotating frame starting with a microscopic Hamiltonian. We show that the two versions of the semi-classical EOMs are in complete agreement up to a phase space coordinate transformation which modifies coordinates and momenta at the first order in $\hbar$. We derive the generalized CVE current which agrees with the literature in both the fermionic and photonic cases and extends the previous studies to an arbitrary spin.

\section{Berry phase}
Let us start with a brief review of the Berry phase for particles of an arbitrary spin by using the path integral formalism, see e.g.~\cite{Nowak:1990wu, Stephanov:2012ki,Yamamoto:2017gla, Yamamoto:2017uul}. The Hamiltonian describing a massless Weyl-type particle of spin $s$ is given by~\cite{Bacry:1975di},
\begin{eqnarray}
\label{hamil}
H=\frac{1}{s}{\bm S}\cdot{\bm p},
\end{eqnarray}
associated with the subsidiary condition
\begin{eqnarray}
\label{cons}
i {\bm S}\times{\bm p}+\frac{1}{s}{\bm S}({\bm S}\cdot{\bm p})-s{\bm p}=0,
\end{eqnarray}
where vector-matrices ${\bm S}$ are $(2s+1)\times(2s+1)$-dimensional and form a representation of $SU(2)$. As usual, the eigenvalues of any component of ${\bm S}$ are $-s, \cdots, s$.

Now we should consider the standard path integral
\begin{eqnarray}
\label{path1}
Z=\int {\cal D}\vec x{\cal D}\vec p {\cal P} e^{iI/\hbar}~~,~~I=\int dt [\dot {\bm x}\cdot{\bm p}-H]\,,
\end{eqnarray}
where $\cal P$ denotes the path ordering. We have to diagonalize the Hamiltonian to evaluate the path integral, to do so one can introduce a set of unitary matrices $V_{p}$ for each ${\bm p}(t)$, such that
\begin{eqnarray}
\label{path1}
V^\dag_{p} H V_{p}=\frac{1}{s}|{\bm p}| S_z,
\end{eqnarray}
where $S_z={\rm diag}(s, \cdots, -s)$. The subsidiary condition projects onto the physical states with only two possible helicities and energies $|{\bm p}|$ and $-|{\bm p}|$. The states with energy $|{\bm p}|$ corresponds to particles while the states with energy $-|{\bm p}|$ corresponds to either antiparticles, if the particles are charged, or unphysical states, if the particles are neutral. We consider only particle states and antiparticle states can be similarly treated. Then, one can rewrite the path integral as
\begin{eqnarray}
\label{path2}
&~&Z=\prod_{1}^{N-1} d{\bm x}_i\prod_{1}^N d{\bm p}_i  {\cal P}\prod_{i=1}^N\notag\\
&~&~~~~~~~~~~~\times\exp{\left[\frac{i}{\hbar}\Delta t\left({\bm p}_i\cdot\frac{{\bm x}_i-{\bm x}_{i-1}}{\Delta t}-\frac{1}{s}{\bm S}\cdot{\bm p}_i\right)\right]}~~~~
\end{eqnarray}
noting that at two adjacent points on the trajectory,
\begin{eqnarray}
\label{path3}
&&\exp{\left(-i\frac{1}{s}{\bm S}\cdot{\bm p}_2\right)}\exp{\left(-i\frac{1}{s}{\bm S}\cdot{\bm p}_1\right)}= V_{p_2}\exp{\left(-i\frac{1}{s}|{\bm p}_2|S_z\right)}V^\dag_{p_2}V_{p_1}\exp{\left(-i\frac{1}{s}|{\bm p}_1|S_z\right)}V^\dag_{p_1}\nonumber\\
&&~~~~~~~~~~~~~~~~=V_{p_2}\exp{\left(-i\frac{1}{s}|{\bm p}_2|S_z\right)}\exp{(-i\hat{{\bm a}}_{p_1}\cdot\dot{{\bm p}}_1\Delta t)}\exp{\left(-i\frac{1}{s}|{\bm p}_1|S_z\right)}V^\dag_{p_1},
\end{eqnarray}
where $\hat{{\bm a}}_{p}=-i V^\dag_{p}{\bm \nabla}_p V_{p}$ and we have assumed $\Delta{\bm p}={\bm p}_2-{\bm p}_1$ to be small in obtaining the expression above. Taking the semi-classical limit during which the off-diagonal components of $\hat{{\bm a}}_{p}$ can be ignored, we find the action which takes the form
\begin{eqnarray}
I_s=\int dt [{\dot {\bm x}}\cdot{\bm p}-\hbar{\bm a}_{p}\cdot\dot{{\bm p}}-\varepsilon_{p}]\,,
\label{action}
\end{eqnarray}
with the chiral dispersion relation $\varepsilon_{p}=|{\bm p}|$ and ${\bm a}_{p}=[\hat{{\bm a}}_{p}]_{ss}$ being the Berry connection. From this derivation one can find that ${\bm a}_p$ results in an emergent ``magnetic'' field in the momentum space\footnote{The detailed calculation of ${\bm b}$ can be found in the Appendix.} ${\bm b}={\bm\nabla}_p\times {\bm a}_p=\pm s\frac{\hat {\bm p}}{|{\bm p}|^2}$ where the sign corresponds to two possible helicities (The derivation above is, however, only for positive helicity).

Further, investigating the analogy between the fermionic case $s=\frac{1}{2}$ and higher spins, it is instructive to highlight the similarity in the  behavior of the general semi-classical action under Lorentz transformations which are known to be modified by the topological phase. An infinitesimal Lorentz boost should leave the semi-classical action invariant if one ignores the topological phase in Eq.~(\ref{action}), see \cite{Skagerstam:1992er, Chen:2014cla, Duval:2014ppa}. The naive infinitesimal Lorentz boost is parameterized by
\begin{eqnarray}
\delta_\beta {\bm x}={\bm \beta} t~,~\delta_\beta t={\bm\beta}\cdot {\bm x}~,~\delta_\beta {\bm p}={\bm\beta}|{\bm p}|
\label{boost}
\end{eqnarray}
and it is straightforward to check that Eq.~(\ref{action}) transforms as
\begin{eqnarray}
\delta_\beta I=-\hbar\int p\beta_i\mathcal{F}_{ij}dp_j
\end{eqnarray}
where $\mathcal{F}_{ij}=\partial_i a_{p,j}-\partial_j a_{p,i}=\ve_{ijk} b_k$ is the Berry field strength. Recovering the form used in \cite{Chen:2014cla}, we can rewrite the variation as
\begin{eqnarray}
\label{variations}
\delta_\beta I=\hbar\int s\frac{{\bm \beta}\times {\bm p}}{|{\bm p}|^2}\cdot\dot {\bm p}~dt\,
\end{eqnarray}
and one can see that the leading correction to the action due to the Berry phase is not invariant if the naive prescription for the boost (\ref{boost}) is used.

The modified prescription can be deduced from the uncanceled part of the variation (\ref{variations}). The new Lorentz transformation takes the form
\begin{eqnarray}
\label{lorentz_new}
\delta^\prime_\beta {\bm x}={\bm\beta} t+s\hbar\frac{{\bm \beta} \times {\bm p}}{|\bp|^2}~,~\delta^\prime_\beta t={\bm\beta}\cdot {\bm x}~,~\delta^\prime_\beta {\bm p}=|\bp| {\bm\beta}~~~
\end{eqnarray}
and corresponds to a symmetry of the action up to the next-to-leading order in $\hbar$.

Despite that the generalization is straightforward due to the analogy between semiclassical actions for chiral massless particles of different spins, let us briefly describe the angular momentum conservation known to be non-trivial in the fermionic case. For an elastic scattering of two particles the angular momentum is conserved. Moreover, in the center of mass frame and with zero impact parameter, the spin and orbital contributions to the angular momentum are zero both in the initial and final states. Following the logic of \cite{Chen:2014cla} we boost along the velocity of an incoming particle. The initial angular momentum is unchanged and stays zero in this setup. Nevertheless, outgoing particles are boosted, their momenta are not collinear anymore, and their spins cannot cancel each other. Note that the momentum and spin are slaved for a chiral particle. On the other hand, the full angular momentum should be zero if the final particles are coming from the same point and the conservation is seemingly violated as noticed in \cite{Chen:2014cla}.

However, the new Lorentz transformation (\ref{lorentz_new}) contains a shift of the trajectories leading to a non-zero orbital contribution to the full angular momentum in the final state which exactly compensates the spin contribution in the boosted frame. Indeed,
\begin{eqnarray}
\label{sidddd}
\Delta {\bm x}=s\hbar\frac{{\bm\beta}\times {\bm p}}{|{\bm p}|^2}~\Rightarrow~{\bm L}_{\text{final}}=s\hbar\frac{{\bm\beta}\times {\bm p}}{|{\bm p}|^2}\times {\bm p}
\end{eqnarray}
while the total spin of the scattered particles is given by the boost of the unit vector along the momentum multiplied by the spin value
\begin{eqnarray}
{\bm S}_{\text{final}}=s\hbar\delta_\beta\frac{{\bm p}}{|{\bm p}|}=s\hbar\left({\bm\beta}-\frac{({\bm \beta}\cdot {\bm p}){\bm p}}{|{\bm p}|^2}\right)
\end{eqnarray}
and one concludes that ${\bm L}_{\text{final}}=-{\bm S}_{\text{final}}$ for any value of $s$. Thus, we have arrived at a generalization of the famous side jump concept for chiral particles with $s>\frac{1}{2}$, see e.g. \cite{Skagerstam:1992er, Chen:2014cla}.

\section{Rotating frame: Intuitive picture}
We have derived the semi-classical action (\ref{action}) for a particle of an arbitrary spin and seen that the Berry phase affects the angular momentum conservation. For our purpose of studying CVE, one has also to take into account non-inertial effects due to the system rotation. However, there is no agreement on the proper modifications of the semi-classical EOMs in the literature, see e.g. \cite{Stephanov:2012ki,Chen:2012ca,Chen:2014cla,Jiang:2015cva,Dayi:2016foz}. To make a connection between various considerations, we rely on two distinct approaches to the problem and show that the final EOMs can be brought to the same form by a phase space coordinate transformation at the leading order in $\hbar$.

Let us start our consideration with the simplest choice of the semi-classical EOMs in a rotating frame which are obtained from the semi-classical EOMs in electromagnetic field by naively replacing the Lorentz force with the Coriolis force~\cite{Stephanov:2012ki}, then
\begin{eqnarray}
&&\dot{{\bm x}}=\frac{\partial \ve}{\partial {\bm p}}+\hbar\dot{{\bm p}}\times {\bm b}~~,~~\dot{{\bm p}}={\bm F}+2|{\bm p}|\dot{{\bm x}}\times {\bm \Omega}\,.
\label{semieom}
\end{eqnarray}
where ${\bm F}$ is an external force, ${\bm\Omega}$ is the angular velocity of the rotating frame, and the last term represents the Coriolis force for a massless particle. The effect of forcing ${\bm F}$ is needed to derive the spin Hall effect \cite{Bliokh:2009ek} along with CVE and it may include both non-inertial effects and a regular forcing. Note that the equations describing a semi-classical trajectory for an arbitrary spin are identical with the fermionic case, up to the value of $s$ hidden in $\ve$ and ${\bm b}$.

It is crucial to note that the energy of a chiral particle should be shifted by a spin-orbit interaction \cite{Chen:2014cla} and, at the leading order in powers of the angular velocity,
\begin{eqnarray}
\ve=|{\bm p}|-\Phi-\hbar s\left(\hat {\bm p} \cdot {\bm \Omega}\right)\,,
\label{edens1}
\end{eqnarray}
where we have introduced a potential $\Phi$ corresponding to the force $\vec F$. That not only modifies the EOMs but also results in an anisotropic distribution function entering the CVE current.

To proceed further one has to express the time derivatives in Eq.~(\ref{semieom}) through non-derivative terms, transforming equations to the form
\begin{eqnarray}
&&\sqrt{G}\dot{{\bm x}}=\frac{\partial \ve}{\partial {\bm p}}+\hbar{\bm F}\times {\bm b}+2\hbar|{\bm p}|~{\bm\Omega}(\hat {\bm p}\cdot {\bm b})\notag\\
&&\sqrt{G}\dot{{\bm p}}={\bm F}+2|{\bm p}|~\hat {\bm p}\times {\bm \Omega}+2\hbar|{\bm p}|~{\bm b}({\bm F}\cdot {\bm \Omega})\,,
\label{semieomsolved}
\end{eqnarray}
where $G=\left(1+2\hbar|{\bm p}|~{\bm b}\cdot{\bm \Omega}\right)^2$ is the determinant of the coefficient matrix in Eq.~(\ref{semieom}). It should be stressed that $\sqrt{G}dx^3 dp^3$ is the invariant phase space measure obeying the Liouville equation~\cite{Sundaram:1999,Xiao:2009rm}. This measure is required to properly treat the axial anomaly at the semi-classical level in the case of massless fermions in external electromagnetic fields, see e.g. \cite{Stephanov:2012ki, Son:2012wh}.

Before moving further it is instructive to take a closer look at the velocity along the trajectory. Expanding up to the linear order in ${\bm \Omega}$, one obtains
\begin{eqnarray}
\dot{{\bm x}}=\hat {\bm p}+s\hbar\frac{{\bm F}\times \hat {\bm p}}{|{\bm p}|^2}+s\hbar\frac{{\bm \Omega}_\perp}{|{\bm p}|}
\label{velocity}
\end{eqnarray}
where $\Omega_\perp^i=\left(\delta^{ij}-\hat{p}^i\hat{p}^j\right)\Omega^j$. The second term on the rhs is a single-particle form of the spin Hall effect for an arbitrary spin $s$ which consists in a shift between trajectories of the two polarizations in the presence of a forcing\footnote{For a recent discussion on spin Hall effect for gravitons, see \cite{Yamamoto:2017gla}}. The last term in Eq.~(\ref{velocity}) is caused by non-inertial effects and can be associated with an interplay between the Coriolis force and the Berry phase. Since this term results in the CVE current in a chiral medium, as we will see shortly, it should be considered as a single-particle realization of this chiral effect\footnote{From (\ref{velocity}) one may note that $|\dot {\bm x}|=1+\mathcal{O}({\bm\Omega}^2)$ for an arbitrary spin in the absence of an external force. It is a natural result since a rigid body rotation is a coordinate transformation and in a relativistic theory one expects massless particles to propagate with the speed of light. It is in contrast with the response to an external magnetic field for fermions \cite{Kharzeev:2016sut} where the velocity gains a contribution only along $\hat {\bm p}$.}. Interestingly, the similarity between the spin Hall effect and CVE at the one-particle level is not stressed in the literature to the best of our knowledge and we highlight that here.

With the semi-classical EOMs in hand, we can write down a semi-classical kinetic equation for the distribution function $f(t,{\bm p}, {\bm x})$ in phase space:
\begin{eqnarray}
\label{ckt1}
\pt_t f +\dot{{\bm x}}\cdot\vec\nabla_x f+\dot{{\bm p}}\cdot\vec\nabla_p f=C[f],
\end{eqnarray}
with $C[f]$ the collision term and $\dot{{\bm x}}$ and $\dot{{\bm p}}$ given by Eq.~(\ref{semieomsolved}). This is the chiral kinetic equation. We now can derive the CVE current in a rotating thermal gas of particles with an arbitrary spin $s$. Since we are interested not only in charged particles it is more convenient to introduce the current as a number current \cite{Stephanov:2012ki, Son:2012wh}
\begin{eqnarray}
{\bm J}=\int_p\sqrt{G}\dot{{\bm x}}f({\bm p}, {\bm x})\,,
\label{current}
\end{eqnarray}
where $f$ should satisfy Eq.~(\ref{ckt1}) and $\int_p=\int \frac{d^3p}{(2\pi)^3}$. The problem is strongly simplified at equilibrium in which $f$ depends only on $\ve$. There are two distinct contributions to the CVE current: the linear-in-${\bm \Omega}$ part of $\sqrt{G}\dot{{\bm x}}$ and the anisotropy of the distribution function due to the energy shift $f\left(|{\bm p}|-s\hbar\left(\hat {\bm p} \cdot {\bm\Omega}\right)\right)$. The full CVE current takes the form
\begin{eqnarray}
{\bm J}=\hbar\int_p\frac{s}{|{\bm p}|}\left(\left({\bm \Omega}+\hat {\bm p}(\hat {\bm p}\cdot {\bm\Omega})\right)f(|{\bm p}|)-{\bm p} \left(\hat {\bm p} \cdot {\bm\Omega}\right)f^\prime(|{\bm p}|)\right)\,.\notag
\end{eqnarray}
Averaging by angles we find that in a rotating medium there is a current along the rotation axis. If both left- and right-handed contributions are taken into account, one finds that in a rotating medium there is a chirality transfer given by
\begin{eqnarray}
{\bm J}_{\chi}=\hbar\frac{s{\bm \Omega}}{\pi^2}\sum_{\pm}\int_0^\infty f_\pm(|\bp|)|\bp|d|\bp|\,.
\label{sCVE}
\end{eqnarray}
In the fermionic case $s=\frac{1}{2}$ and the chirality transfer corresponds to the axial current, $f(|\bp|)$ is the Fermi-Dirac distribution function. One could find  that the resulting value of the CVE current coincides with other derivations and is given by ${\bm J}_{\chi}=\hbar\left(\frac{\mu_+^2+\mu_-^2}{4\pi^2}+\frac{T^2}{6}\right){\bm\Omega}$, where $\mu_\pm$ are chemical potentials for right- and left-handed fermions, see e.g. \cite{Vilenkin:1980zv, Son:2009tf, Sadofyev:2010is, Landsteiner:2011cp}. One should note that Eq.~(\ref{sCVE}) is general and gives the chirality transfer in a rotating system of massless particles with an arbitrary $s$ including CVE for photons \cite{Avkhadiev:2017fxj,Yamamoto:2017uul, zuyizin}.

It is worth mentioning that in many setups the semi-classical EOMs take the form (\ref{semieom}) with a modified ``spin-orbit'' coupling in $\ve$ leading, in principle, to a similar transport at equilibrium, see e.g. \cite{Cortijo:2016wnf, Hayata:2017tbr}. However, the energy shift due to the ``spin-orbit'' interaction can naively modify the current and velocity along the trajectory. Interestingly, it appears that the energy shifts in the distribution function and in the $\sqrt{G}\dot{\bm x}$ cancel each other, $\int_p (\pt\ve/\pt\bm p) f(\ve)=\int_p\pt \tilde{f}(\ve)/\pt{\bm p}=0$ with $\tilde{f}'(x)=f(x)$. This fact, noticed previously for the response of $s=\frac{1}{2}$ fermions to magnetic field \cite{Kharzeev:2016sut}, appears to be general and also explains the agreement between derivations based on \eq{semieom}  despite whether the ``spin-orbit'' correction is taken into account or omitted.

\section{Rotating frame: First principles}
We have seen how the CVE current (\ref{sCVE}) can be derived for an arbitrary spin using the semi-classical EOMs (\ref{semieom}) and chiral kinetic theory. While most relevant non-inertial effects are included intuitively in Eq.~(\ref{semieom}), this approach is far from being satisfactory and it would be instructive to derive the EOMs from first principles. One may note that \eq{sCVE} is in agreement with the common answer for the CVE in the fermionic case which was extensively studied in the literature, see e.g. \cite{Vilenkin:1980zv, Son:2009tf, Sadofyev:2010is, Landsteiner:2011cp, Stephanov:2012ki, Chen:2014cla}. Thus, while the semi-classical EOMs  (\ref{semieom}) can be incomplete, we expect the CVE current to be the same. In this section we will consider an explicit derivation of the semi-classical EOMs in a rotating frame seeking for a more accurate description along a similar strategy as in~\cite{Huang:2015mga}.

We start with a single ``right-handed" free massless particle of spin $s$ in a rotating frame and use $\hbar$ as an expansion parameter. The Hamiltonian (\ref{hamil}) is modified due to the coupling between the full angular momentum and the angular velocity \cite{Hehl:1990nf,Chen:2015hfc}:
\begin{eqnarray}
\label{hamilrot}
H=\frac{1}{s}{\bm S}\cdot{\bm p}-\Phi-{\bm \O}\cdot\lb{\bm x}\times{\bm p}+ \hbar\vec {\bm S}\rb,
\end{eqnarray}
where the dynamics of $\bm S$ is constrained by the same subsidiary condition (\ref{cons}). We note here that the Hamiltonian (\ref{hamilrot}) is the one measured by a rotating observer while the momentum $\bp$ in Eq.~(\ref{hamilrot}) is still define in the inertial frame, see discussion in Ref.~\cite{Liu:2018xip}. The Heisenberg EOMs corresponding to this Hamiltonian are given by
\begin{eqnarray}
\label{eomx}
\dot{\bm x}&=&\frac{1}{s}{\bm S}-{\bm \O}\times{\bm x}\notag\\
\label{eomp}
\dot{\bm p}&=&{\bm \nabla}_x \Phi +{\bm p}\times{\bm \O}\notag\\
\label{eoms}
\dot{\bm S}&=&\frac{1}{s\hbar}{\bm p}\times{\bm S}-{\bm\O}\times{\bm S}\,.
\end{eqnarray}
One can solve for the spin degrees of freedom order by order in $\hbar$:
\begin{eqnarray}
\label{expan1}
{\bm S}={\bm S}_0 + \hbar{\bm S}_1 + \cdots\,,
\end{eqnarray}
deriving in this way the semi-classical EOMs for the orbital variables $\bm x$ and $\bm p$. From \eq{eoms}, we get ${\bm S}_0=\l\hat{\bp}$ with $\l=-s,\cdots,s$. For a given $\l$, the vector $\bm S$ must satisfy ${\bm S}^2=\l^2$ leading to a constraint ${\bm S}_0\cdot{\bm S}_1=0$. This condition combined with \eq{eoms} gives
\begin{eqnarray}
{\bm S}_1 &=& \l\frac{s}{|\bp|}\lb\dot{\hat{\bm p}}\times\hat{\bm p}+\hat{\bm p}\times(\hat{\bm p}\times{\bm \O})\rb.
\end{eqnarray}
Imposing the subsidiary condition (\ref{cons}), one finds, as expected, that it projects onto the two possible physical states with $\l=\pm s$. This is consistent with the fact that a massless particle of an arbitrary spin has only two independent degrees of freedom.

Substituting $\bm S_0$ and $\bm S_1$ into \eq{eomx}, we obtain the time evolution of $\bm x$ and $\bm p$,
\begin{eqnarray}
\label{eom2}
\dot{\bm x}&=&\hat{\bm p}-{\bm \O}\times{\bm x}+\hbar{\bm \nabla}_x \Phi\times{\bm b}\non
&=&{\bm \nabla}_p\ve+\hbar\dot{\bm p}\times{\bm b}\notag\\
\dot{\bm p}&=&{\bm \nabla}_x \Phi+{\bm p}\times{\bm\O}\non
&=&-{\bm \nabla}_x\ve\,,
\end{eqnarray}
where the dispersion relation corresponding to the Hamiltonian (\ref{hamilrot}) is given by
\begin{eqnarray}
\label{enrgdens}
\ve= |{\bm p}|-\Phi-{\bm \O}\cdot\lb{\bm x}\times{\bm p}\rb-s\hbar{\bm \O}\cdot \hat{\bm p}\,.
\end{eqnarray}
One can readily see that \eqs{eom2}{enrgdens} differ from the EOMs and the dispersion relation used in the previous section. Surprisingly, the EOMs (\ref{eom2}) are seemingly independent $\hbar$ if the external force is turned off.  Note that for neutral particles, e.g. photons, the antiparticle state with $\lambda=-s$ is meaningless, so we keep only $\lambda=s$ case which corresponds to the right circularized photon. To get the left circularized photon, one has to change the sign in \eq{hamil}.

The chiral kinetic equation take the same form as Eq.~(\ref{ckt1}) but with $\dot{\bm x}$ and $\dot{\bm p}$ now given by Eq.~(\ref{eom2}).
We are interested in the CVE current which can be found from a similar expression as (\ref{current}) in the previous section. However, one can readily find that there are two main differences comparing to the previous section at equilibrium: 1) the measure of the phase space integral is trivial; 2) the distribution function explicitly depends on the coordinates through the dispersion relation. If we rely on the same definition of the current we will obtain only $1/3$ of the expected CVE response. This is due to the lack of the magnetization current associated with the side-jump effect, which should be taken into account in order to ensure the Lorentz symmetry \cite{Chen:2014cla},
\begin{eqnarray}
{\bm J} &=&\int_p\left(\dot{\bm x}-\hbar |{\bm p}|{\bb}\times{\bm \nabla}_x\right) f({\bm p}, {\bm x}).
\end{eqnarray}
Note that in principle one had to consider a similar contribution in Eq.~(\ref{current}). However, the dispersion relation (\ref{edens1}) involves no coordinate dependence in a uniformly rotating medium and this term can be safely omitted.

At equilibrium $f=f(\ve)$ the current linear in $\hbar$ reads
\begin{eqnarray}
{\bm J}_\hbar &=&c\hbar{\bm \nabla}_x \Phi\times\int_p {\bm b} \ls f(\ve)-|{\bm p}|f'(\ve)\rs\non&&-s\hbar\int_p\ls(c\hat{\bm p}-{\bm\O}\times{\bm x}){\bm\O}\cdot\hat{\bm p}-c\hat{\bm p}\times(\hat{\bm p}\times{\bm \O})
\rs f'(\ve)\,.\notag
\end{eqnarray}
One may note that the side-jump effect contributes to the integrated current both for the spin Hall effect and for the CVE. Considering the CVE current at the rotation axis and keeping only terms linear in $\bm \O$ we find
\begin{eqnarray}
{\bm J}_\hbar &=&-cs\hbar\int_p\ls \hat{\bm p}\,\hat{\bm p}\cdot{\bm \O}-\hat{\bm p}\times(\hat{\bm p}\times{\bm \O})
\rs f'(\ve),\non
&=&-s\hbar{\bm \O}\int_p\ls \frac{1}{3}+\frac{2}{3}\rs f'(\ve)\,,
\end{eqnarray}
or, integrating by parts and summing over chiralities,
\begin{eqnarray}
{\bm J}_{\chi}=\hbar\frac{s{\bm \O}}{\pi^2}\sum_{\pm}\int f_{\pm}(|\bp|)|\bp|d|\bp|\,
\label{sCVE2}
\end{eqnarray}
which reproduces \eq{sCVE}. In next section, we will show that the two descriptions based on the EOMs (\ref{semieomsolved}) and (\ref{eom2}) are, in fact, related by a phase space transformation.

\section{Coordinate transformation}
As was mentioned above, the agreement between (\ref{sCVE}) and (\ref{sCVE2}) is expected from general arguments. However, it is interesting to find the underlying reasoning for this peculiar fact. Since the phase space measure is different in two considerations it is natural to assume that there is a coordinate transformation bringing the two sets of EOMs to the same form and, thus, also links the two chiral kinetic theories. Here we will seek for redefined notions of coordinate and momentum to see how the two pictures relate with each other. Let us call the new momentum and coordinate $\bk$ and $\by$ to avoid a confusion with notations used before, taking a simple ansatz
\begin{eqnarray}
\label{tran1}
&\by= \bx + \ba (\bx,\bp)~~,~~\bk=\bp+\bA(\bx,\bp)\,.
\end{eqnarray}
We can re-write \eq{eom2} as a matrix relation
\begin{eqnarray}
\begin{pmatrix}
\dot{x}_i\\
\dot{p}_i
\end{pmatrix}
=\begin{pmatrix}
\z_{x_ix_j} & \z_{x_ip_j}\\
\z_{p_ix_j} & \z_{p_ip_j}
\end{pmatrix}
\begin{pmatrix}
\frac{\pt \ve}{\pt x_j}\\
\frac{\pt \ve}{\pt p_j}
\end{pmatrix},
\end{eqnarray}
where $\z$ is the minus inverse of the symplectic form. 
Comparing with \eq{eom2}, we find that its components are given by
\begin{eqnarray}
&&\z_{x_ix_j}=-\hbar\mathcal{F}_{ij}=-\hbar\e_{ijk}b_k,\non
&&\z_{x_ip_j}=\d_{ij},~~\z_{p_ix_j}=-\d_{ij},~~\z_{p_ip_j}=0\,.
\end{eqnarray}
One can write a similar relation for the new coordinates
\begin{eqnarray}
\label{eommod}
\begin{pmatrix}
\dot{y}_i\\
\dot{k}_i
\end{pmatrix}
=\begin{pmatrix}
\d_{ij}+\frac{\pt a_i}{\pt x_j} & \frac{\pt a_i}{\pt p_j}\\
\frac{\pt A_i}{\pt x_j} & \d_{ij}+\frac{\pt A_i}{\pt p_j}
\end{pmatrix}
\begin{pmatrix}
\z_{x_jx_l} & \z_{x_jp_l}\\
\z_{p_jx_l} & \z_{p_jp_l}
\end{pmatrix}
\begin{pmatrix}
\d_{ls}+\frac{\pt a_s}{\pt x_l} & \frac{\pt A_s}{\pt x_l}\\
\frac{\pt a_s}{\pt p_l} & \d_{ls}+\frac{\pt A_s}{\pt p_l}
\end{pmatrix}
\begin{pmatrix}
\frac{\pt \ve}{\pt y_s}\big|_k\\
\frac{\pt \ve}{\pt k_s}\big|_y
\end{pmatrix}.
\end{eqnarray}
Note that the invariant measure of the phase space coordinates $(\by,\bk)$ is given by the transformation Jacobian. As we have seen, the difference between the two dispersion relations is crucial and we eliminate the explicit $\bx$ dependence in $\ve$ first. This can be achieved by choosing
\begin{eqnarray}
\label{Apot}
\bA=|\bp|\bx\times\bO,
\end{eqnarray}
because in such case we can express $\ve$ as,
\begin{eqnarray}
\label{vare}
\ve=|\bk|-\Phi-s\hbar\bO\cdot\hat{\bk}+\mathcal{O}(\bO^2),
\end{eqnarray}
neglecting $\mathcal{O}(\bO^2)$ terms. Next, we assume that $\Phi$ is a function of $\by$ solely and $\ba$ is $\mathcal{O}(\hbar)$ and $\mathcal{O}(\O)$ because without quantum effects and rotation there is no need to have position shift. Finally, we are interested in the region near $\bx\sim 0$ so that we can neglect terms explicitly $\propto \bx$ in the final EOMs (but not in (\ref{enrgdens})). In fact, the region far from the rotating axis requires a more careful discussion, see e.g.~\cite{Vilenkin:1980zv}.

Within the mentioned approximation, we have
\begin{eqnarray}
\label{eommod2}
\begin{pmatrix}
\dot{y}_i\\
\dot{k}_i
\end{pmatrix}
&=&\begin{pmatrix}
\z_{x_ix_s}& \d_{is}+\frac{\pt a_i}{\pt x_s}+\z_{x_i x_l}\frac{\pt A_s}{\pt x_l}\\
\frac{\pt A_i}{\pt x_j}\z_{x_j x_s}-\d_{is}-\frac{\pt a_s}{\pt x_i}& \frac{\pt A_i}{\pt x_s}-\frac{\pt A_s}{\pt x_i}
\end{pmatrix}
\begin{pmatrix}
\frac{\pt \ve}{\pt y_s}\big|_k\\
\frac{\pt \ve}{\pt k_s}\big|_y
\end{pmatrix}+\mathcal{O}(\hbar^2, \bO^2, \bx)\,,
\end{eqnarray}
and the invariant measure of the phase space $(\by,\bk)$ is reduced to
\begin{eqnarray}
\label{meas}
\sqrt{G}=1-\frac{\pt a_i}{\pt x_i}.
\end{eqnarray}
Finally, choosing $\ba$ to be
\begin{eqnarray}
\ba=\hbar\,|\bp|\,\bb\times(\bO\times\bx),
\end{eqnarray}
we find that \eq{eommod2} and \eq{meas} take the desired form as given in Eq.~(\ref{semieomsolved}) (in the considered limit)
\begin{eqnarray}
\label{eomnew}
&&\sqrt{G}\dot{\by}=\frac{\partial \ve}{\partial\bk}+\hbar{\bm F}\times {\bm b}+2\hbar|\bk|~{\bm\Omega}(\hat\bk\cdot {\bm b})\notag\\
&&\sqrt{G}\dot{\bk}={\bm F}+2|\bk|~\hat\bk \times {\bm \Omega}+2\hbar|\bk|~{\bm b}({\bm F}\cdot {\bm \Omega}),
\end{eqnarray}
where $\sqrt{G}=1+2\hbar |\bk|\bO\cdot\bb$ and $\ve$ is given by Eq.~(\ref{vare}). So we find that the two sets of EOMs presented in the previous sections are in complete agreement up to a phase space coordinate transformation and lead to the same CVE current.

These equations can be also compared with the EOMs for right-handed fermions ($s=1/2$) in electromagnetic field expanded to the leading order in powers of fields,
\begin{eqnarray}
\label{eomb}
&&\sqrt{G}\dot{\bx}=\frac{\partial \ve_B}{\partial\bk}+\hbar{\bm E}\times {\bm b}+\hbar\,{\bm B}(\hat\bk\cdot {\bm b})\notag\\
&&\sqrt{G}\dot{\bk}={\bm E}+\hat\bk \times {\bm B}+\hbar\,{\bm b}({\bm E}\cdot {\bm B})
\end{eqnarray}
with $\bm k$ being the kinetic momentum of the particle, $\varepsilon_B=|\bk|-\Phi-\hbar{\bm B}\cdot\hat{\bk}/(2|\bk|)$ being the dispersion in magnetic field \cite{Son:2012zy} (where $\Phi$ should be understood as Coulomb potential), and the phase space measure $\sqrt{G}=1+\hbar (\bB\cdot\bb)$. A replacement $\bm E \leftrightarrow\bm F, \bm B\leftrightarrow 2|\bm k|\vec\Omega$ links the sets (\ref{eomnew}) and (\ref{eomb}) as well as the corresponding phase space measures. (But the correspondence between $\ve$ and $\ve_B$ is different due to the influence of Land\'e $g$-factor which is $2$ rather than $1$ for spin-$1/2$ fermions.)

Let us comment about the physical meaning of the transformations (\ref{tran1}). For this purpose, we consider a local, instantaneous, Lorentz boost which transforms the frame with coordinate $\bx$ and momentum $\bp$ to a frame with coordinate $\bm y$ and momentum $\bm k$. The boost velocity is nothing but $\vec\beta=-\vec\O\times\bx$ (Note that we are considering the region near the rotating axis so that $|\vec\beta|$ is small). Thus we find that our coordinate transformation has the following sense: 1) Before the transformation, the energy is defined in the rotating frame corresponding to a Killing vector $\pt_0$ in terms of the momentum $\bp$ measured in the inertial frame. After the transformation, the same energy is expressed in terms of the momentum $\bk$ measured in the rotating frame. Thus, Eq.~(\ref{tran1}) is a local Lorentz transformation of the phase space from the inertial frame to the rotating frame. Recently, the same conclusion was also drawn from a more involved method based on the Wigner function defined in curved spacetime in Ref.~\cite{Liu:2018xip}. 2) The momentum of the particle at $\bx$ in the inertial frame transforms to the corresponding momentum in the rotating frame according to Eq.~(\ref{lorentz_new}), $\d'_{\beta}\bp=|\bp|\vec\beta$, which is exactly the field $\vec A$ in Eq.~(\ref{Apot}). The field $\vec A$ can also be considered as an effective gauge field with the field strength $\vec B_{\rm eff}=\vec\nabla\times\vec A=-2|\bp|\vec\O$ giving a relativistic extension of the well-known Larmor theorem for massive particles. 3) The position transformation $\vec a$ is nothing but the side jump given in Eq.~(\ref{sidddd}). This effect in the transformed coordinates does not appear in the expression for the CVE current (as the magnetization contribution) but is accounted exactly by the position transformation. We note that $\d'_{\b}\bx$ given in Eq.~(\ref{lorentz_new}) gives an additional term $-(\bO\times\bx)t$ which, however, does not change the EOMs given in Eq.~(\ref{eomnew}) at $\mathcal{O}(\bx^0)$. The time is the same before and after the boost, $\d'_\b t=\vec\b\cdot\bx=0$.

\section{Summary}
In summary, we have developed a semi-classical framework for massless particles of arbitrary spin $s\geq 1/2$ in rotating frame. Our framework can apply to both single-particle problem and many-body problem. The former is described by a set of semi-classical single-particle equations of motion (EOMs) and the latter is described by the chiral kinetic theory. We consider two different formulations of the EOMs as well as the chiral kinetic theory and show that they are connected by a coordinate transformation in the phase space and are, thus, equivalent. Such a coordinate transformation is further recognized as a special Lorentz boost containing the so-called side-jump effect. In particular, in one of the two formulations, the EOMs are in close analogy to those in electromagnetic field assembling a relativistic Larmor relation between rotation and magnetic effects. Our finding clarifies the superficial discrepancies in the literature between different formulations of the chiral kinetic theory for rotating systems.

The spin Hall effect for an arbitrary spin appears naturally as a Berry phase effect in our formalism. We show that both the spin Hall effect and its cousin, the chiral vortical effect, are a general feature of massless particles with an arbitrary spin. We give explicit expression for the  chirality current in a general rotating chiral medium. The higher-spin realizations of chiral vortical effect can be in principle observed in various setups including table-top experiments on quantum optics.

\section*{Acknowledgments}
We would like to thank Daekyoung Kang who participated at early stages of this study. We thank Yuchen Liu, Kazuya Mameda, Misha Stephanov for helpful discussions. The work of AS is supported through the LANL/LDRD Program and by RFBR Grant 18-02-40056. XGH is supported by the Young 1000 Talents Program of China, NSFC through Grants No. 11535012 and No. 11675041.

\appendix
\section{Calculation of the Berry curvature}
In this Appendix, we calculate the Berry curvature ${\bm b}$ introduced below Eq.~(\ref{action}), following closely Ref.~\cite{Berry:1984jv}. We start with Eq.~(\ref{path1}) in which the $(2s+1)\times (2s+1)$-dimensional unitary matrix $V_p$ can be written as $V_p=(C_s, C_{s-1}, \cdots, C_{-s})$, where $C_l(\bp)$ ($l=-s, \cdots, s$) is the eigenvector of $H$ with eigenvalue $ l|\bp|/s$. The Berry connection ${\bm a}_\bp=[\hat{\bm a}_p]_{ss}=-iC_s^\dag\vec\nabla_p C_s$ for positive helicity. The corresponding Berry curvature is
\begin{eqnarray}
{\bm b}&=&-i\vec\nabla_pC_s^\dag\times\vec\nabla_p C_s\nonumber\\
&=&i\sum_{l=-s}^{s-1}C_s^\dag\vec\nabla_p C_l\times C_l^\dag\vec\nabla_p C_s,
\end{eqnarray}
where we have used the orthogonality and completeness properties of $\{C_l\}$, $C^\dag_{l'} C_l=\delta_{ll'}$ and $\sum_{l=-s}^s C_l C_l^\dag=1$. Using the property $ C^\dag_{l'} S_i C_l=(l'-l)|\bp| C_{l'}^\dag\pt_{p_i}C_l$ for $l'\neq l$ which can be obtained by differentiating the eigen-equation $H C_l=(l|\bp|/s)C_l$, we have
\begin{eqnarray}
\label{app:b}
{\bm b}&=&-i\sum_{l=-s}^{s-1}\frac{1}{(l-s)^2|\bp|^2}C_s^\dag{\bm S} C_l\times C_l^\dag{\bm S} C_s.
\end{eqnarray}
To proceed, we temporarily choose the axes so that the $z$-axis is along the direction of $\bp$ and employ the relations,
\begin{eqnarray}
(S_x\pm iS_y)C_l&=&[s(s+1)-l(l\pm1)]^{1/2}C_{l\pm1},\\
S_z C_l&=& l C_l.
\end{eqnarray}
It is clear that only the term with $l=s-1$ in the summand of Eq.~(\ref{app:b}) can contribute to ${\bm b}$ and only $b_z$ is nonzero:
\begin{eqnarray}
b_z&=&-i\frac{1}{|\bp|^2}\left(C_s^\dag S_x C_{s-1} C_{s-1}^\dag S_y C_s-C_s^\dag S_y C_{s-1} C_{s-1}^\dag S_x C_s\right)\nonumber\\
&=&\frac{s}{|\bp|^2}.
\end{eqnarray}
Recovering to a general coordinates, we obtain ${\bm b}=s\hat{\bp}/|\bp|^2$.

\bibliographystyle{bibstyle}
\bibliography{jumps_new}

\end{document}